# Spatial Filter with Volume Gratings for High-peak-power Multistage Laser Amplifiers

Yi-zhou TAN*[a], Yi-sheng YANG[b], Guang-wei ZHENG[b], Ben-jian SHEN[b], Heng-yue PAN[c], LIU Li[b]

[a]College of Mechatronic Engineering and Automation, [b]College of Science, [c]College of Computer,

National University of Defense Technology, Changsha, Hunan, China, 410073

## ABSTRACT

The regular spatial filters comprised of lens and pinhole are essential component in high power laser systems, such as lasers for inertial confinement fusion, nonlinear optical technology and directed-energy weapon[1,2]. On the other hand the pinhole is treated as a bottleneck of high power laser due to harmful plasma created by the focusing beam. In this paper we present a spatial filter based on angular selectivity of Bragg diffraction grating to avoid the harmful focusing effect in the traditional pinhole filter. A spatial filter consisted of volume phase gratings in two-pass amplifier cavity were reported. Two-dimensional filter was proposed by using single Pi-phase-shifted Bragg grating, numerical simulation results shown that its angular spectrum bandwidth can be less than 160urad. The angular selectivity of photo-thermo-refractive glass and RUGATE film filters, construction stability, thermal stability and the effects of misalignments of gratings on the diffraction efficiencies under high-pulse-energy laser operating condition are discussed.

## 1. INTRODUCTION

A typical spatial filter comprises a first lens, a pinhole and a second lens behind the pinhole[1,2]. In this paper a new type of spatial filter based on angular selectivity of Bragg diffraction grating is proposed in which neither the lens nor the pinhole were used for high-pulse-energy lasers. Two types of spatial filters were investigated in following passages: (a) spatial filter in two-pass laser amplifier cavity comprised of a pair of volume Bragg gratings, (b) spatial filter with single Pi-phase-shifted Bragg grating as horizontal and vertical filtering elements simultaneously.

### 1.1 The requirements of the narrow-pass band filter for high power laser beam

For practically application high power laser beam should be as clean as possible. The lens-pinhole combination acts as a low-pass filter to control the nonlinear growth of spatial noise during propagation of high-power laser beams [1]. Initial confinement fusion experiments utilized pulses typically of 100-psec FWHM. For these short pulses spatial filters with f/10 optics and 300-Mimp inholes were found to have energy transmissions of 98% or greater [2]. The high-pulse-energy lasers are typically composed of a chain of laser amplifiers with a diameter increasing from input to output. Transport telescopes in these lasers are required to match the beam diameter with the amplifier apertures. The quality of lenses and telescope alignment governs the laser beam quality, and determines the stability and reliability of the laser. For above-mentioned reasons, spatial filters may be viewed as one of the key elements of intense laser systems.

Pinhole spatial filters have been used for more than 35 years to "clean up" the spatial structure of a laser beam. On the other hand the combination of lens-pinhole is a bottleneck of high power laser. Usually, intensities in the central regions of the beam can reach $10^{17}$ W $cm^2$; pinhole is sufficiently small with the result that the focusing beam striking the edge of the pinhole creates a plasma. The plasma can distort or block the transmitted pulse, or lead to back reflections [1, 3]. In order to avoid the shortcoming of pinhole filters we present volume Bragg grating spatial filter as substitute of the traditional pinhole filter in high power solid-state disk laser amplifier.

### 1.2 Comparison volume phase grating spatial filter with the traditional pinhole spatial filter

Test setup for demonstration of the spatial filtering effect is shown in Figure 1 and Figure 2 respectively.

---

* tanyizhou@126.com; phone 086-28-85211710

In Fig.1 pinhole spatial filter uses the principles of Fourier optics to alter the spatial structure of a laser beam, removing aberrations in the beam Diameter of pinhole is 1.0mm and focus of Fourier lens is 450mm. In Fig.2 the filtering element is VBG [Volume Bragg Gratings], its parameter shows in Table 1. This filter operates with an unfocused, because it is not necessary to produce Fourier transform for laser beam cleanup.

The modulator generates a deformed beam contained high frequency amplitude modulation noise. 1-D intensity distribution of the deformed beam in our demonstration experiment is shown in Fig. 2 (right).

Table 1 Characteristics of the Phase VBG (Photo-Thermal-Refractive glass)

| Performances | Values | Performances | Values |
| --- | --- | --- | --- |
| Grating Period | 3.06 (um) | Grating Thickness | 2.88 (mm) |
| Tilt | -0.02 (deg) | Grating Dimensions | 15×15 (mm) |
| Refractive index | 1.485 | Modulation of the refractive index | $1.82 \times 10^{-4}$ |

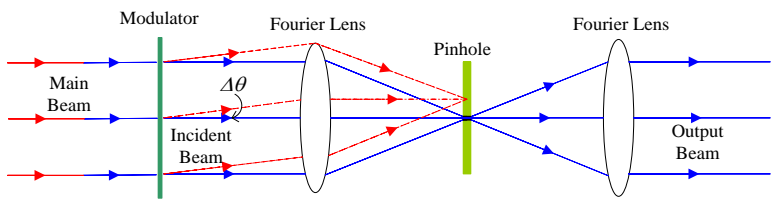

Fig.1 Experimental configuration of pinhole filter

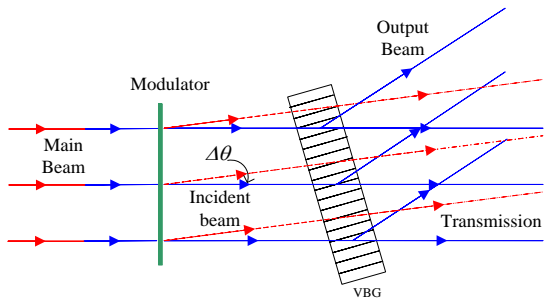 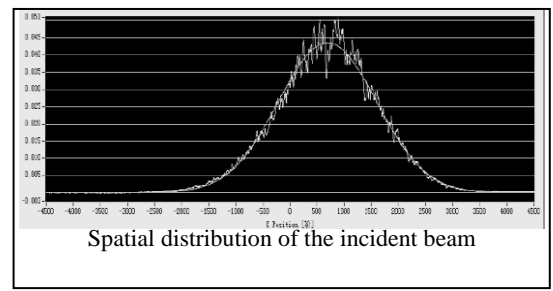

Spatial distribution of the incident beam

Fig.2 Experimental configuration of VBG filter

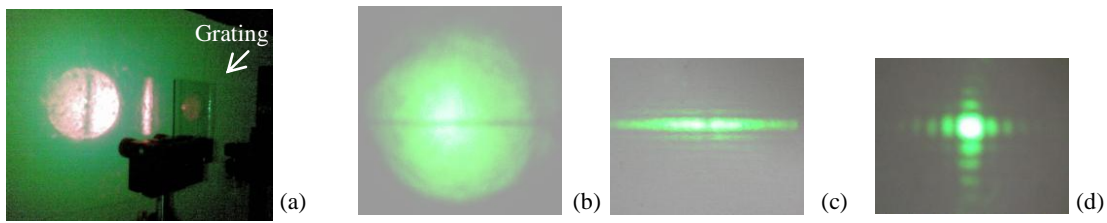

(a)  (b)  (c)  (d)

Fig.3 Two-Dimensional spatial filtering process by volume phase gratings
(a) Experiment setup and its 1-D spatial filtering result ($\lambda_b$ =633nm), (b) Laser main spot ($\lambda_b$ =512nm),
(c) Horizontal spatial filtering stripe, (d) 2-D Spatial filtering spot (low pass filtering result)

Figure 3-a shows 1-D spatial filtering result of He-Ne laser beam. Two-dimensional filtering experiments were carried out by using a pair of VBG elements [4, 5]. The pictures from Fig.3-b to Fig.3-d show horizontal axis filtering and vertical axis filtering results in series.

The VBG filter share same modulator with pinhole filter. Figure 4 shows that both VBG filter and pinhole filter can filter out high-frequency components (noise) of the spatial spectrum. The lower angular spectrum components remained in the output beam.

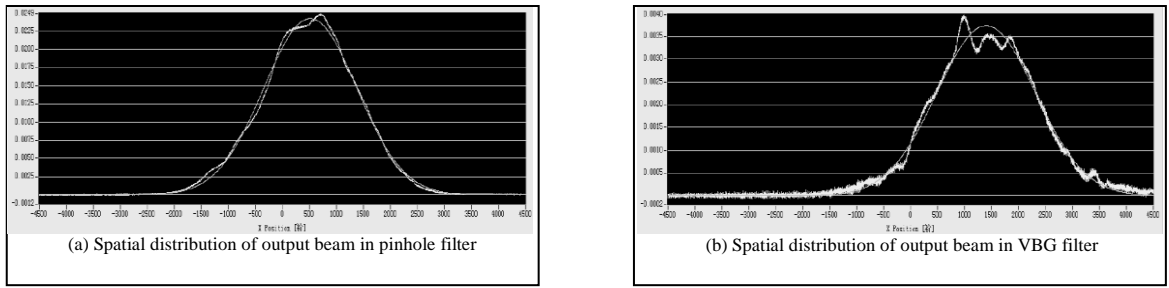

Fig.4 The intensity profiles after spatial low-pass filtering (measured by Laser Beam Profiler BP109-UV)

## 2. THE ANGULAR SELECTIVITY OF VOLUME BRAGG GRATINGS

For a volume phase grating with a sinusoidal uniform volume structure produced by refractive index modulation, the angle of incidence $\theta_b$ for a perfectly phase matched Bragg diffraction is given by

$$\cos\theta_b = \lambda/2n\Lambda \qquad (1)$$

Here we assume that the grating fringes are perpendicular to the surface of the slab medium. $\Lambda$ is grating periodicity, $n$ is the average index of refraction of the grating, $\lambda$ is the vacuum wavelength of incident light.

According to [6], relationship between spectral and angular parameters could be obtained from differential form of Bragg-matching condition (1)

$$\Delta\theta_b/\Delta\lambda = \Lambda/2nF_{\pi/2}, \quad \text{where} \quad F_{\pi/2} = \sqrt{1-(\lambda\Lambda/2n)^2} \qquad (2)$$

When the laser incident angle is different than the Bragg angle, the relationship between diffraction efficiency $\eta_\theta$ and incident angle deviation $\Delta\theta$ is can approximated as typical *sinc* function

$$\eta_\theta = \eta_B \operatorname{sinc}^2(\theta-\theta_B)/\Delta\theta \qquad (3)$$

As a example, for the typical angle of incidence ( 45 deg) used in the Bragg diffraction filter, the beam deviation is approximately 90 deg, according to formula (3), for a filter with the wavelength sensitivity $\Delta\lambda/\lambda \approx 10^{-3}$, the angular sensitivity $\Delta\theta/\theta$ would also be 1-mrad (or more). One can see that the angular selectivity is good, but not as low as predicted, it not suitable for high power laser which need beam quality near Diffraction-limited.

In the following section three methods were proposed to solve this problem: (1) Let laser beam pass the same grating through back and forth optical paths, (2) Combing two VBG to form a Pi-phase-shifted Bragg grating, or (3) Use the Pi-phase-shift RUGATE film as filtering element.

## 3. CONFIGURATION OF SPATIAL FILTERS IN HIGH-ENERGY PULSED LASER

### 3.1 Two- Dimensional spatial filter in two-pass (multi-pass or multistage) laser amplifier cavity

This type of spatial filter is comprised of a pair of volume Bragg gratings with grating vectors perpendicular to each other [4, 5]. In Fig.5-a and Fig.5-b the optical path have a similar shape, but the volume Bragg gratings are operating in different mode: in transmitting mode, the main beam ( Bragg wavelength) pass through the filter with a minimal loss, and selected wavelength is deflected to go away from the laser amplifier cavity; in reflecting mode, the reflected wavelength at the Bragg angle is main beam.

There is a combination conventional laser gain medium with a 2-D spatial filter; a mirror was mounted at the right end of cavity in order to forming a back and forth optical path (Figure 5). Bragg diffraction effect of same grating appeared twice for one laser beam in two-pass laser amplifier cavity. The VBG diffraction efficiency curves ( Fig.5-b) show that the low-pass filtering effects of second pass is good than the first one. An explanation for improving of beam smoothing effects is that the thickness of grating expands nearly 50%.

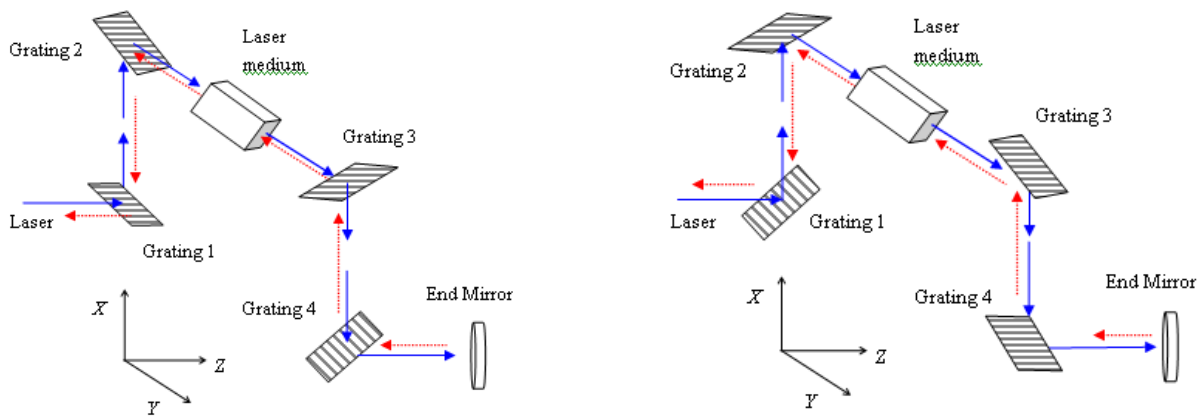

Fig.5-a Combination of spatial filters with laser medium in two-pass Laser amplifier cavity
*The grating operating mode: transmitting grating mode (Left), reflection grating mode (Right)

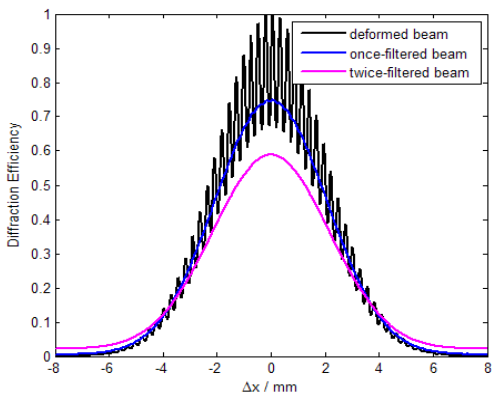

Fig.5-b Beam smoothing effects for volume phase grating in back and forth operating mode in setup of Figure5-a

## 3.2 With Pi-phase-shifted Bragg gratings to carry out 2-D spatial filtering

### I. Arrangement of gratings in the filter

We design two types of two-Dimensional filter with Pi-phase-shifted Bragg grating, as shown in Fig.6.

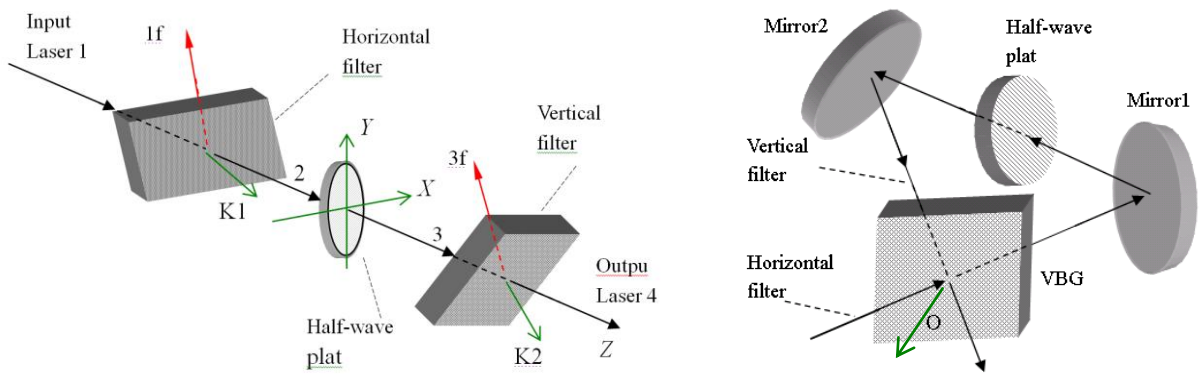

Fig.6 Two-Dimensional spatial filter consist of Pi-phase-shifted Bragg gratings
Lift: a pair of gratings in series with its grating vectors perpendicular to each other;
Right: Single grating act as Horizontal and Vertical filtering elements simultaneously

In Fig.6 (Lift), the spatial filter comprises horizontal filtering grating, half-wave plate and vertical filtering grating. Its optical axis is a straight line. In Fig.6 (Right), the spatial filter comprises grating, half-wave plate and two mirrors. Its optical axis is a triangle. We could treat the triangle path as a holding shape of the straight line path in Fig.6 (lift) by merging the two gratings into one (In practice, let one beam pass through same grating twice).

The notable feature of filter containing one piece of VBG, compared to a regular filter containing two pieces of VBG, is that the single grating act as Horizontal and Vertical filtering elements simultaneously. The filtering process in triangle path is as following: in the first step the incident laser beam is transmitted though the grating under the condition of that both the laser ray and the grating's normal vector are laying in the horizontal plane, in the second step the laser beam incident the grating again under the condition of that both the laser ray and the grating's normal vector are laying in the Vertical plane. Two reflective mirrors are necessary to change filter operating state from Vertical mode to Horizontal mode by changing the beam propagate directions along the triangle path. The first filtering process and the second filtering process do not disturb each other in the non-slanted transmission grating.

**II. Pi-phase-shifted Bragg grating for ultra-narrow bandwidth spatial filtering**

Some high-pulse-energy laser system, such as lasers for inertial confinement fusion, directed-energy weapon, nonlinear optical technology and so on, need low-pass bandwidth less than 100 urad The $\pi$-phase-shifted Bragg grating is a useful component for ultra-narrow bandwidth filter [7, 8]. The main feature of Pi-phase-shifted Bragg grating, compared to a regular single piece of VBG, is that it is coherent combination of two separate pieces of VBG with a Pi-phase-shifted middle layer. One could perhaps treat this type of components as a synthesis of symmetrical Bragg gratings and Fabry–Peŕot (F-P) cavity.

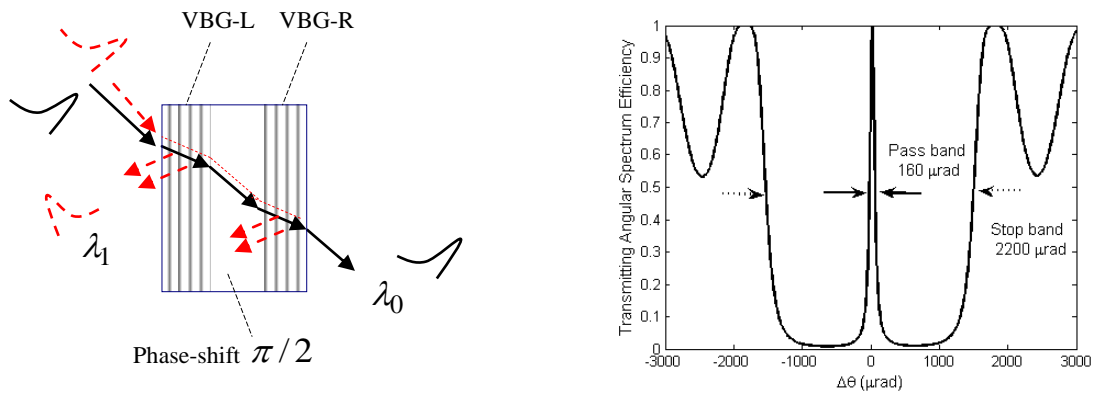

Fig.7 Pi-phase-shifted Bragg grating and its angular spectrum selectivity curve

A Pi-phase-shifted grating have sandwich structure (Fig.7-Left), as an example, which grating parameters are in Table 2. Its one-dimensional spatial filtering for the deformed Gaussian laser beam is simulated by chain-matrix analysis. The results show that when the phase-shift of middle layer is odd times of $\pi/2$, the $\theta$-domain transmission bandwidth can be less than 160urad (Figure7-Right).

Table 2 Characteristics of the Pi-phase-shifted VBG (wavelength at 1064nm)

| Performances | Values | Performances | Values |
|---|---|---|---|
| Grating Period | 400 (nm) | Grating Thickness (Lift, Right) | 500 (um) |
| Incident angle | 0.48 (rad) | Thickness of middle layer | 250nm |
| Mean relative permittivity | 2.25 | Modulation of the refractive index | 0.0036 |

Both the Photo-Thermal-Refractive glass Bragg grating and RUGATE film have same design of refractive-index profile for a narrow bandpass filter: a typical index profile consists of a layer of mean index, equal in thickness to one (or odd times) half-period of the rugate index oscillation, with symmetrical sinusoidal periods on either side.
It is noticeable that sinusoidal index variation occurs naturally within hologram gratings [6]. It is also possible to make interference filters in which the index of refraction varies in a smooth fashion as a function of depth by film deposition technology. Fabrication of RUGATR thin films requires stringent control of the refractive index of the film during deposition [7, 8].

## 4. CONCLUSION

The angular selectivity of regular Volume Bragg Grating is 1mrad order of magnitude, it is meaning that single piece of VBG can't satisfy the requirements of ultra-narrow bandwidth filter. Aforementioned design and experiment results indicate that two methods are effective for narrowing the angular selectivity by one order of magnitude at least:

- Laser beam propagate through same grating twice (Double the Grating's effective thickness);

- Use coherent combination of two pieces of VBG or RUGATE film to form a $\pi$-phase-shift filtering components.

Due to their high angle and wavelength sensitivity, and no heating and optical distortions in the exposed areas at power density up to 100 kW/cm$^2$ (for Photo-thermo-refractive glass), Volume Bragg Gratings suitable for high power laser filter. Its drawbacks are:

- Works properly for a fixed wavelength only;

- For typical volume Bragg gratings the relative angular selectivity and wavelength selectivity are similar, so it is difficult to design an ultra-narrow bandwidth spatial filter for ultra-short pulse laser (Broadening effects in frequency-domain).

The yaw, pitch, and roll of the grating (which are due to the construction stability about the z, y, and x axes, respectively), together with thermal stability, are disadvantageous phenomenon on grating's diffraction efficiencies. In order to substitute VBG filter for traditional pinhole spatial filter, an ultra-narrow bandwidth VBG filter need to decrease the harmful effects of misalignments of grating axes, especially pitch misalignment.

The drift of Bragg wavelength due to changes of temperature is needed to compensate by incorporating a passive thermally actuated supporting parts of grating [5].

**ACKNOWLEDGMENT**
This work was supported by NSAF Foundation of National Natural Science Foundation of China and Chinese Academy of Engineering Physics (CAEP) under the Project No.10676038, also acknowledges partial support from The 1st Innovation Experiment & Inquiry Learning Program for university students (The Education Department of Hunan Province, China）.